\documentclass[12pt]{article}
\usepackage{amsmath}
\usepackage{amssymb}
\usepackage{graphics}
\usepackage{epsf,graphicx,rotating,color}
\usepackage{bm,bbm}
\usepackage{amsfonts}
\begin{document}
\title{{\bf One dimensional relativistic free particle in a quadratic dissipative medium}}
\author{G.V. L\'opez, G.C. Montes, and J.G.T. Zanudo\\\\Departamento de F\'{\i}sica, Universidad de Guadalajara,\\
Blvd. Marcelino Garc\'{\i}a Barrag\'an 1421, esq. Calzada Ol\'{\i}mpica,\\
44420 Guadalajara,  Jalisco, M\'exico}
\date{October 2009}
\maketitle
\begin{abstract}
The deduction of a constant of motion, a Lagrangian, and a Hamiltonian for  relativistic 
particle moving in a dissipative medium characterized by a force which depends on the square of
the velocity of the particle is done. It is shown that meanwhile the trajectories in the space ($x,v$), defined by the constant of motion, look as one might expected, the trajectories in the space ($x,p$), defined by the Hamiltonian, have a odd behavior. 
\end{abstract}
\newpage
\section{\bf  Introduction}
\noindent
It is well known that the Lagrangian and Hamiltonian approaches for some non dissipative and some
dissipative systems have some problems  [1-6].  One of these problems consist in the possibility of having two different Hamiltonian to the same classical system [7], implying that one will have two different quantization for this system. Another problem consist that for some dissipative non relativistic systems, like a free particle moving in a dissipative medium characterized by a force which depends on the square of the velocity of the particle, the  trajectories on the space ($x,p$) have an odd behavior. However, the trajectories on the space ($x,v$), defined by the constant of motion,  have a good expected
behavior [8].\\\\
In this work , the study of this former problem is extended to the relativistic motion of the particle. The constant of motion, the Lagrangian, and the Hamiltonian are deduced consistently , and it is shown that the behavior of the trajectories of the particle in the phase space ($x,p$) are odd when the Hamiltonian approach is used. However, the trajectories in the space ($x,v$), when the constant of motion is used, behave as one could expected .\\\\
\section{\bf Constant of motion, Lagrangian and Hamiltonian}
\noindent
The one-dimensional motion of a relativistic particle  of mass $"m"$ at rest which is moving with a velocity $\dot x=dx/dt$ in a dissipative medium characterized by a force which depends on the square of this velocity is described by the equation
\begin{equation}
\frac{d}{dt}(m\gamma \dot x)=-\alpha\dot x^2\ ,
\end{equation}
where $\alpha$ is the dissipative parameter, $c$ is the speed of light, and $\gamma$ is the relativistic factor, $\gamma=(1-\dot x^2/c^2)^{-1/2}$. Actually, Eq. (1) represents a dissipative system for $\dot x\ge 0$, otherwise it represents an anti-dissipative system. Therefore, only the case $\dot x\ge 0$ will be considered below.  This system can be written as the following dynamical system
\begin{equation}
\dot x=v\ ,\quad\quad\quad
\dot v=-\frac{\alpha v^2}{m}\left(1-v^2/c^2\right)^{3/2}\ .
\end{equation}
A constant of motion for this system is a function  $K=K(x,v)$ such that it satisfies the following partial differential equation of first order [9]
\begin{equation}
v\frac{\partial K}{\partial x}-\frac{\alpha v^2}{m}\left(1-v^2/c^2\right)^{3/2}\frac{\partial K}{\partial v}=0\ .
\end{equation}
The general solution of this equation [10] is given by $K=G(C)$, where $G$ is an arbitrary function of the characteristic curve $C$,
\begin{equation}
C=\frac{\alpha}{m}x+\frac{1}{\sqrt{1-v^2/c^2}}-\ln{\left(\frac{1+\sqrt{1-v^2/c^2}}{v/c}\right)}\ .
\end{equation}
By choosing $K=mc^2C$, a constant of motion is gotten with energy units,
\begin{equation}\label{5}
K=\alpha c^2 x+ \frac{mc^2}{\sqrt{1-v^2/c^2}}-mc^2\ln{\left(\frac{1+\sqrt{1-v^2/c^2}}{v/c}\right)}\ .
\end{equation}
The Lagrangian of the system can be  consistently deduced from the known expression 
\begin{equation}\label{6}
L(x,v)=v\int\frac{K(x,v)}{v^2}~dv
\end{equation}
which establishes  the relation between the Lagrangian and the constant of motion of the system [11]. Using this expression it follows that
\begin{equation}\label{7}
L=-\alpha c^2 x-2mc^2\sqrt{1-v^2/c^2}+mc^2\ln{\left(\frac{1+\sqrt{1-v^2/c^2}}{v/c}\right)}\ .
\end{equation}
The generalized linear momentum, $p=\partial L/\partial v$, is given by
\begin{equation}\label{8}
p=mc\frac{2v^2/c^2-1}{(v/c)\sqrt{1-v^2/c^2}}\ .
\end{equation}
The plot of this expression and the plot of the usual relativistic free linear momentum expression ($p=mv/\sqrt{1-v^2/c^2}~$) are shown in Fig. 1, where one sees that for Eq. (8)  there is not  a one to one relation between the velocity $v$ and the generalized linear momentum $p$ of Eq. (8). 
\begin{figure}[H]\label{Fig. 1}
 \begin{center}
 \includegraphics[width=8.5cm,height=6cm]{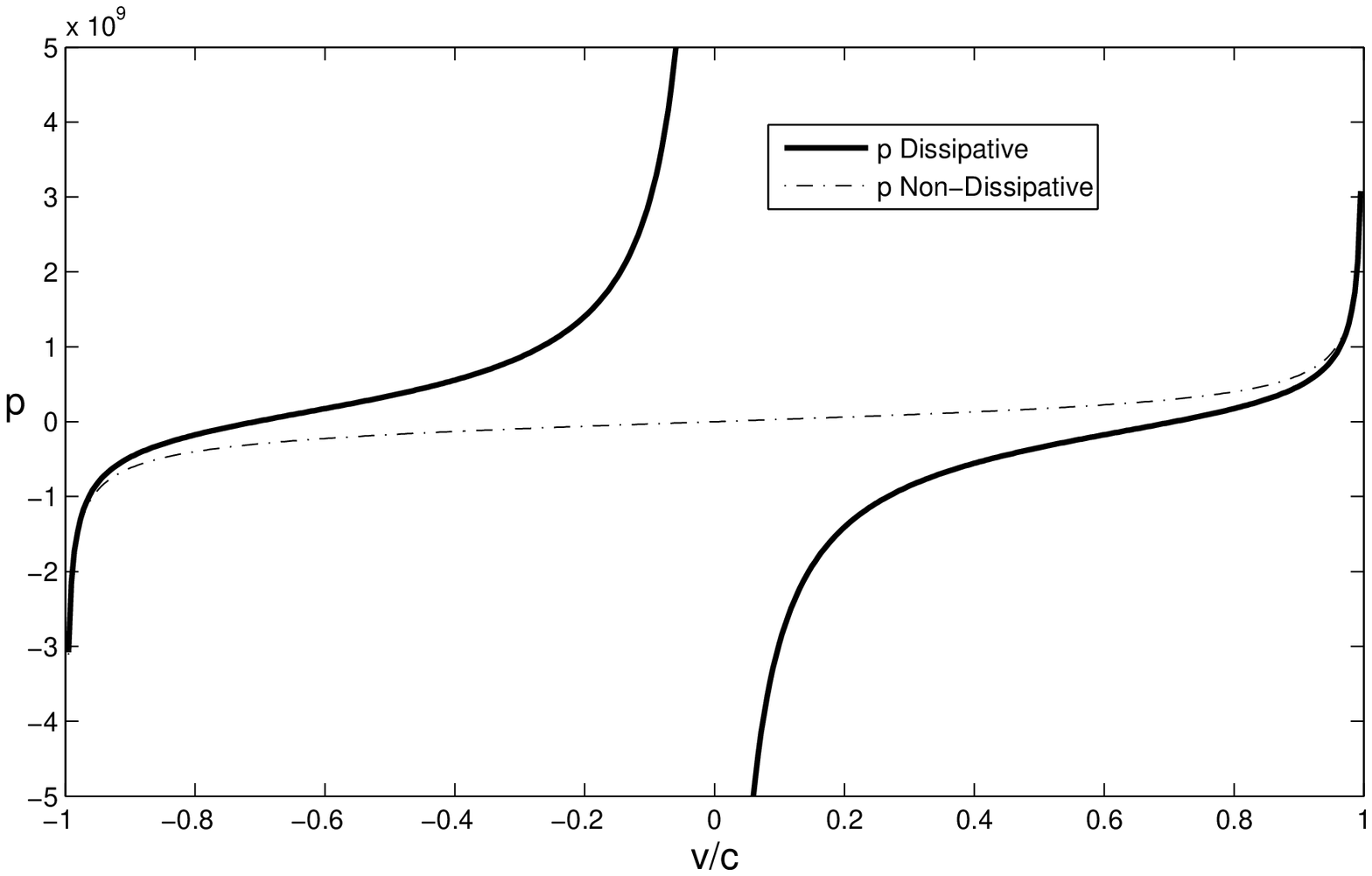}
 \end{center}
 \caption{Relation between the generalized linear momentum and velocity}
\end{figure}
\noindent
The inverse relation of Eq. (8) is shown on Fig. 2, which is given analytically by
\begin{subequations}\label{9}
\begin{equation}
\left(\frac{v}{c}\right)^2_{-}=\frac{1}{2}-\frac{1}{2}\frac{|p|}{\sqrt{p^2+4m^2c^2}}
\qquad\hbox{if}\qquad 0<\frac{v}{c}\leq\frac{1}{\sqrt{2}}
\end{equation}
and
\begin{equation}
\left(\frac{v}{c}\right)^2_{+}=\frac{1}{2}+\frac{1}{2}\frac{|p|}{\sqrt{p^2+4m^2c^2}}
\qquad\hbox{if}\qquad \frac{1}{\sqrt{2}}\leq\frac{v}{c}<1
\end{equation}
\end{subequations}
\begin{figure}[H]\label{Fig. 2}
 \begin{center}
 \includegraphics[width=8.5cm,height=6cm]{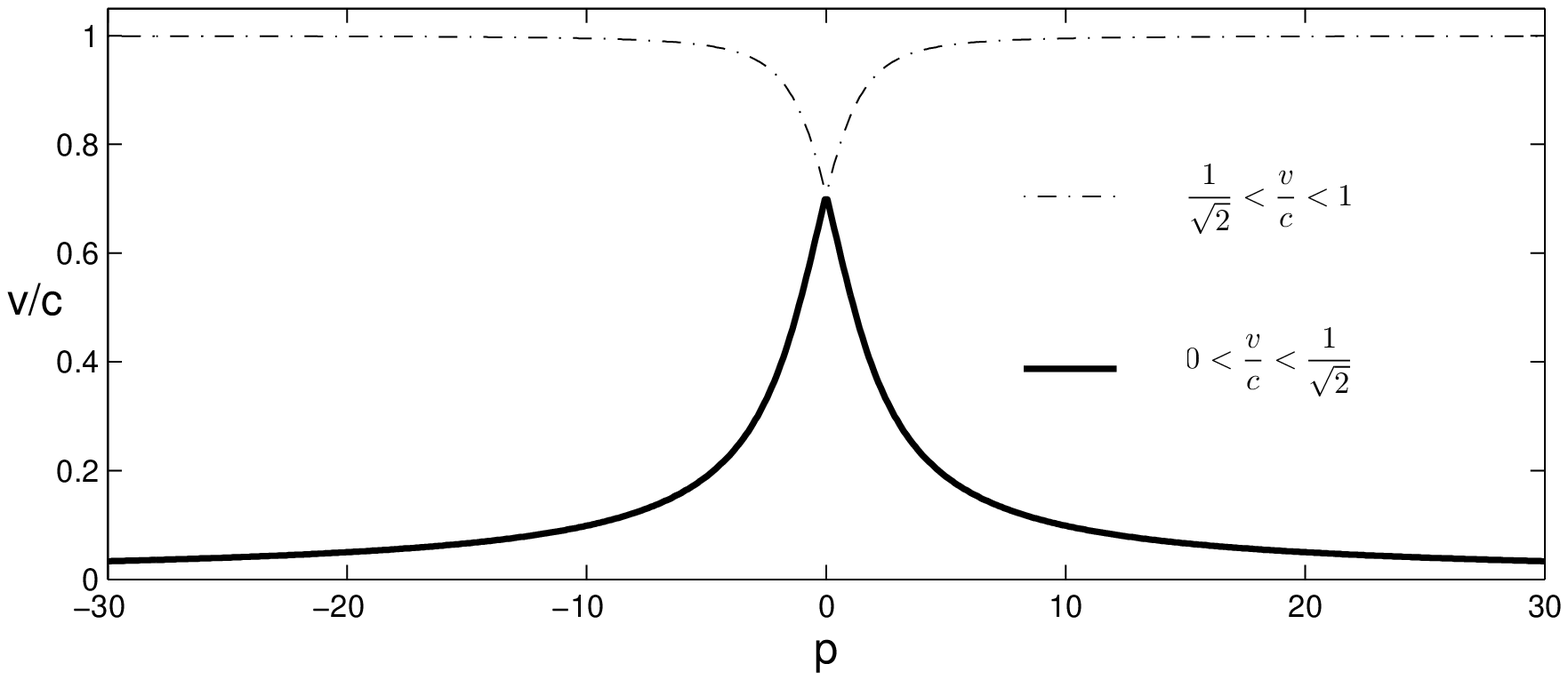}
 \end{center}
 \caption{ Inverse relation between the generalized linear momentum and velocity}
\end{figure}
\noindent
These expressions define respectively the Hamiltonians $H^{(-)}$ and $H^{(+)}$ as
\begin{subequations}\label{10}
\begin{equation}
H^{(-)}=\alpha c^{2}x+\frac{\sqrt{2}m_{0}c^2}{\left[1+\frac{|p|}{\sqrt{p^2+4m^2c^2}}\right]^{1/2}}-m_{0}c^2\ln\left\{\frac{1+\frac{1}{\sqrt{2}}\left[1+\frac{|p|}{\sqrt{p^2+4m^2c^2}}\right]^{1/2}}{\frac{1}{\sqrt{2}}\left[1-\frac{|p|}{\sqrt{p^2+4m^2c^2}}\right]^{1/2}}\right\}
\end{equation}
and
\begin{equation}
H^{(+)}=\alpha c^{2}x+\frac{\sqrt{2}m_{0}c^2}{\left[1-\frac{|p|}{\sqrt{p^2+4m^2c^2}}\right]^{1/2}}-m_{0}c^2\ln\left\{\frac{1+\frac{1}{\sqrt{2}}\left[1-\frac{|p|}{\sqrt{p^2+4m^2c^2}}\right]^{1/2}}{\frac{1}{\sqrt{2}}\left[1+\frac{|p|}{\sqrt{p^2+4m^2c^2}}\right]^{1/2}}\right\}
\end{equation}
\end{subequations}

\section{\bf Trajectories}
\noindent
Using the initial conditions $x=0$, $m=1$ and $v/c=0.7$, the constant of motion (5) is determined and  the trajectories on the space ($x,v$) can be calculated. Fig. 3 shows these trajectories for several values of the parameter $\alpha$. As one can see, the falling down of these trajectories and the way they are falling as the parameter $\alpha$ increases represent the behavior that one can expected for a dissipative medium. Now, given these same initial conditions,  the initial generalized linear momentum is calculated from expression (8). One uses the expression (10a) to determinate the value of this Hamiltonian and to calculated the trajectories in the space ($x,p$). These trajectories can be seen in Fig. 4.  As one can see, these trajectories have an odd behavior since $|p|$ go to infinity as the particle is slowing down, but this was already expected from the same expression for the generalized linear momentum, Eq. (8).\\\\
\begin{figure}[H]\label{Fig. 3}
 \begin{center}
 \includegraphics[width=8.5cm,height=6cm]{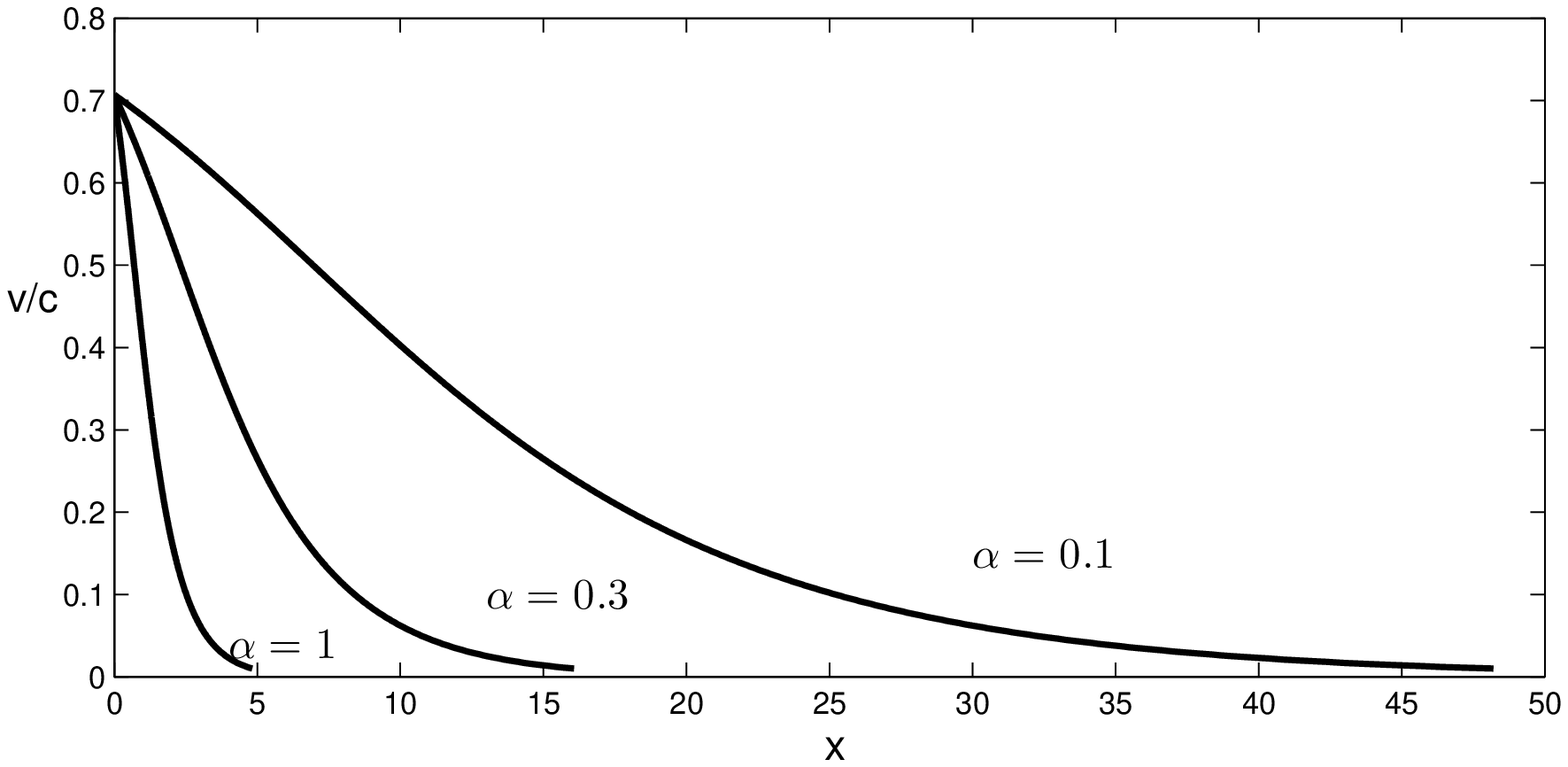}
 \end{center}
 \caption{ Trajectories in the ($x,v$) space, defined by the constant of motion}
\end{figure}
\begin{figure}[H]\label{Fig. 4}
 \begin{center}
 \includegraphics[width=8.5cm,height=6cm]{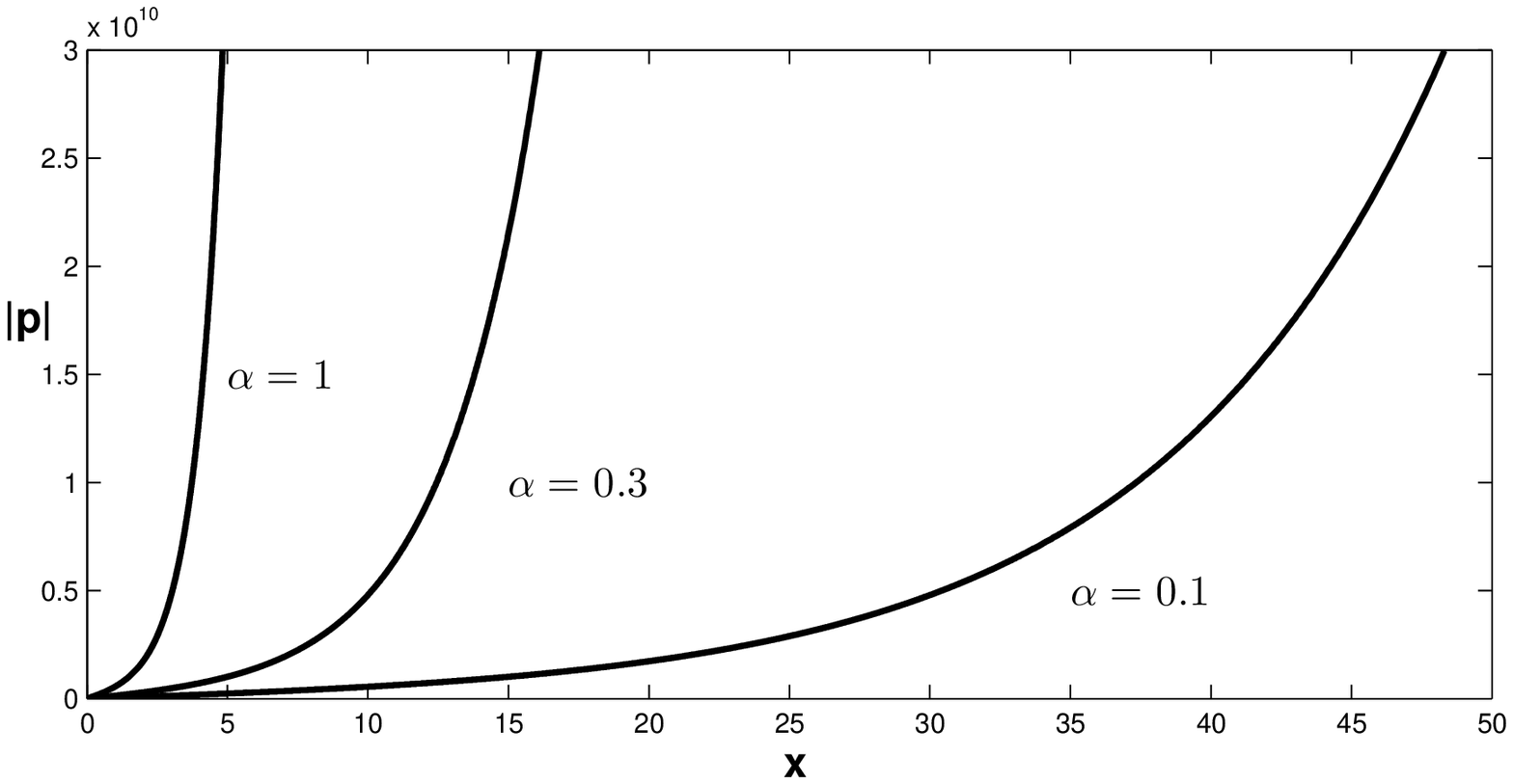}
 \end{center}
 \caption{ Trajectories in the ($x,p$) space, defined by the Hamiltonian}
\end{figure}
\newpage
\section{Conclusion}
\noindent
We have constructed consistently a constant of motion, Lagrangian, and Hamiltonian for a relativistic particle moving in a dissipative medium, characterized  by a force which depends on the square velocity of the particle. The trajectories in the space ($x,v$), defined by the constant of motion, behave as we can expect. However, the trajectories in the space ($x,p$), defined by the Hamiltonian, behave oddly and totally anti-intuitively. This suggest that the Hamiltonian approach may be useless to see the qualitative behavior of a relativistic particle in a dissipative medium.\\\\   

\section{Bibliography}
1. V.P. Dodonov, V.I. Man'ko, and V.D. Skarzhinsky,  Hadronic Journal, {\bf 4} (1981) 1734.\\
2. P.Havas, Act. Phys. Austr., {\bf 38} (1973) 145.\\
3. S. Okubo, Phys. Rev. D, {\bf 22} (1980) 919.\\
4. R. Glauber and V.I. Man'ko, Sov. Phys. JEPT, {\bf 60} (1984) 450.\\
5. G. L\'opez, Ann. Phys. {\bf 251}, no. 2(1996) 372.\\
6. G. L\'opez, Int. J. Theo. Phys., {\bf 37}, no.5 (1998) 1617.\\
7. G. L\'opez, X.E. L\'opez and G. Gonz\'alez, Int. J. Theo. Phys., {\bf 46}, no. 1 (2007) 149.\\
8. G.V. L\'opez, arXiv:0901.4792 (2009).\\
9 . G. L\'opez, {\it Partial Differential Equations of First Order and Their Applications to\\ 
\qquad Physic},  World Scientific, 1999.\\
10. F. John, {\it Partial Differential Equations}, Springer-Verlag, N.Y. 1974.\\
11. J.A. Kobussen, Act. Phys. Austr.,{\bf 51} (1979) 193.\\
12. C. Leubner, Phys. Rev. A, {\bf 86} (1987) 9.\\
13. C.C. Yan, Amer. J. Phys., {\bf 49} (1981) 296.\\
14. See reference 5.

\end{document}